\documentclass[3p,times,twocolumn]{elsarticle}

\journal{Simulation Modelling Practice and Theory}

\def\BibTeX{{\rm B\kern-.05em{\sc i\kern-.025em b}\kern-.08em T\kern-.1667em\lower.7ex\hbox{E}\kern-.125emX}}

\usepackage{times}
\usepackage{gensymb}
\usepackage{tabularx}
\usepackage{lipsum}
\usepackage{array}
\usepackage{booktabs}
\usepackage{listings}
 
\usepackage{enumerate}

\usepackage{multirow}

 \usepackage{graphicx}
 \usepackage{amssymb,amsmath}

\usepackage[caption]{subfig}

\lstdefinestyle{lststyle}{
 captionpos=b, 
 tabsize=2,
 basicstyle=\linespread{0.9}\footnotesize\ttfamily,
}
 
\begin{document}

\begin{frontmatter}

\title{ Using Cloud and Fog Computing for Large Scale IoT-based Urban Sound Classification }

\author[mymainaddress]{Marc Jayson Baucas}
\author[mymainaddress]{Petros Spachos\corref{mycorrespondingauthor}}
\cortext[mycorrespondingauthor]{Corresponding author}
\ead{petros@uoguelph.ca}

\address[mymainaddress]{School of Engineering, University of Guelph, Guelph, Ontario, Canada}

\begin{abstract}
The Internet of Things (IoT) has become the forefront of bridging different technologies together. It brings rise to online computational services that make mundane tasks convenient. However, the volume of devices connecting to the network started to increase. In turn, services that thrived on centralized storage are being strained and overloaded.  As applications and software advances, processing and computational power become a concern to technology companies. With data risks and large numbers of connected devices, cloud computing has become outdated. Devices are forced to commit unnecessary expenses to stay relevant in the market due to the increase in software complexity.  This need for change resulted in the introduction of edge computing.  Edge computing distributes the computational strain between the server and the devices. This contribution allows the cloud to accommodate more users and devices are no longer in need to make significant changes to their design every so often. Many real-time applications have evolved to require high amounts of processing power to execute. For example, sound classification comes with massive computational needs due to its affiliation with neural networks and deep learning. This paper aims to create a feasible and deployable real-time sound classification system. There were three configurations tested in this paper. The results of our experiments show that cloud computing and edge computing alone cannot cater to a technological market that is exponentially growing in size and complexity.  However,  the same results show promise in finding optimal configurations in terms of a combination of end device power consumption, application runtime and server latency to systems instead of focusing on a single model. Overall, it is better to take into consideration the strengths and weaknesses of each computing architecture. In finding a reasonable configuration balance, lower power consumption in end devices and lesser computational strain on cloud servers is a must. 
 
\end{abstract}


\end{frontmatter}

\section{Introduction}
The Internet of Things is a network of heterogeneous devices that are connected to the Internet by wireless access technologies such as Wi-Fi and Bluetooth~\cite{iot-def, Al-Fuqaha}. It provides many online services that are currently being used by the masses. Many have thought to use this network as a means to store data wirelessly to improve convenience for its users. The idea was to allow any user access to his or her data anywhere that has an Internet connection.  The resulting shift in data storage gave rise to services such as; Google Drive, and Dropbox. Every user who had a smart device (i.e. phone, laptop, etc.) gets an online storage space from their providers. 

With the rise of cloud storage, many users have incorporated it within their IoT networks. Most providers have added other services aside from cloud storage to their users. Applications are now available online without having users install them, which created cloud computing. Cloud computing is a way of conducting all forms of data processing remotely in a provided cloud server. Most applications use these servers to implement large scale algorithms such as; natural language translation, image processing, and sound classification \cite{iot-cloud}. As a result, cloud usage removes most of the processing strains on user's devices. However, this makes the process too centralized. This strain is dumped on the cloud instead.  A bottleneck is created that limits the effectiveness of the service. The software and hardware architecture of the server now dictates the scope and limitations of the service. 

The reliance of users on the cloud has been proven to be risky. Security, accountability, and accessibility are some aspects that can be risky to cloud service users~\cite{Takabi}. Relying too much on the cloud to handle all the data processes could be detrimental. It moves all the focus and user's data to the server. If these servers have a security breach or malfunction, the data becomes corrupted. However, pushing all the computation onto the end devices increases its average power consumption. This increase could become detrimental to its lifespan. It also increases the runtime of an application. This increase is due to the significant difference between the processing capabilities of a server and an end device. 

Applications and software that are used by the general public today have increased in complexity. It has reached a point where technology companies are required to make changes often to stay relevant in the market. Having too many computations on one side of the network spectrum can be too costly to both the users and the companies that manage them. These issues in power consumption, runtime, and latency does not introduce edge or fog computing to IoT~\cite{fog-app} as a standalone solution. However, we get a better system by having it in combination with the already existing cloud computing architecture. It can balance the data traffic that strained the cloud servers and reduced the power consumption that limited the end devices. Edge or Fog computing creates a more decentralized means of offloading data processing while Cloud computing maintains a distinct balance in control by not overwhelming the edges. 

One of the computation heavy applications mentioned earlier was sound classification. Sound classification is a type of large scale classification that can be used by cities as a smart city application \cite{road-smart}. However, urban sound classification was selected since it is a more specific subset. This type of implementation deploys a large number of end devices to a city to sample the sound within each respective section. The sound collected is then classified for specific patterns and events that are happening within each area. If each device was to send their sound data continuously, then it could increase their average power consumption which could decrease the lifespan of the system. Also, more processes mean a longer runtime for an application which could result in unreasonable wait times for data to reach the server. This wait time could reduce the integrity of incoming data as it no longer indicates a real-time representation of the sound within the area that the system is classifying. Also, if most of the processing occurs in the cloud, then the server could be overwhelmed and become unresponsive due to the data traffic. 

This paper investigates the effectiveness of using a balanced combination of fog computing and cloud computing for urban sound classification by creating a simulation of the sound classification system. By focusing on the trade-off between end device power consumption, application runtime, and server-side latency, we aim to create a configuration that can make the proposed sensing system scalable. We use a combination of tests and simulations that measure power consumption, runtime, and latency to prove that we can deploy an urban sound sensing platform for IoT networks. 

The rest of this paper is as follows: An in-depth discussion of the different aspects of the proposed system in Section \ref{bground}. Section \ref{method} describes the experimental setup. Next, Section \ref{eval} notes all the conducted experiments and their corresponding observations. Finally, Section \ref{conc} are conclusions to these observations. 

\section{Background} \label{bground}

\begin{figure}
    \centering
    \includegraphics[width=\columnwidth]{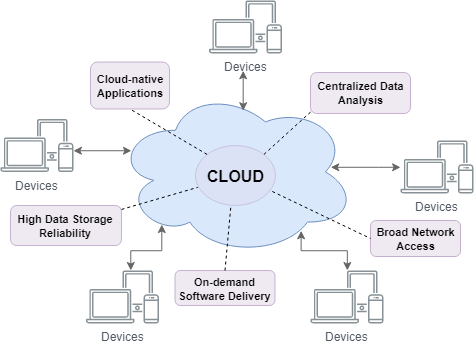}
    \caption{Cloud-based IoT network.}
    \label{cloud-fig}
\end{figure}

\subsection{Cloud Computing}
Cloud computing is a model that enables convenient, on-demand network access to a shared pool of computing resources \cite{cloud-sec}. This shared pool is made possible through cloud storage technology. Cloud storage technology provides users with a means to store, retrieve, or back up their data through an online storage \cite{cloud-ben}. It makes this possible with the use of cluster applications, network technology, distributed file systems, and many more online resources. Partnered with IoT, cloud computing becomes a revolutionary tool for devices that rely on online servers for data storage. 

As technologies progressed, network applications have started demanding more resources and required more processing power. As a result, traditional servers began to have difficulties with the increase in the cost of operation and maintenance.  Cloud computing enters in as a solution by introducing parallel processing. Parallel processing allows a remote server to offload tasks into subroutines systematically. Figure \ref{cloud-fig} shows a diagram that visualizes a cloud-based IoT network and some of its known features \cite{iot-cloud, cloud-ben}. 

However, even though cloud computing has notable strengths over traditional systems. It also comes with some bottlenecks and risks \cite{cloud-sec, cloud-edge-comp}:

\begin{enumerate}[i.]
    \item \textit{Server overloading} - In cloud computing, most devices have their uses reduced to only sending their data to the cloud servers while waiting for the results of the computation. However, as the number of users connected to a cloud increases, data starts being received by servers at higher volumes and velocities. This sudden increase contributes to the data traffic, which results in a bottleneck. Due to data congestion, large-scale servers are significantly slowed down by high latency. High latencies will render a server unresponsive. As a result, the potential of integrating any form of real-time application is reduced by this bottleneck. 
    
    \item \textit{Scalability issues due to centralization} - In a centralized cloud infrastructure, capacity defines the number of users a service can handle. Managing the data traffic within a server will become more tedious if the network funnels all the resources into one centralized unit. A server that can only handle a limited number of devices cannot be scaled up. As a result, when scaling up a cloud-based architecture, its capacity will always be its limiting factor. 

\end{enumerate}

\begin{figure}
    \centering
    \includegraphics[width=1\columnwidth]{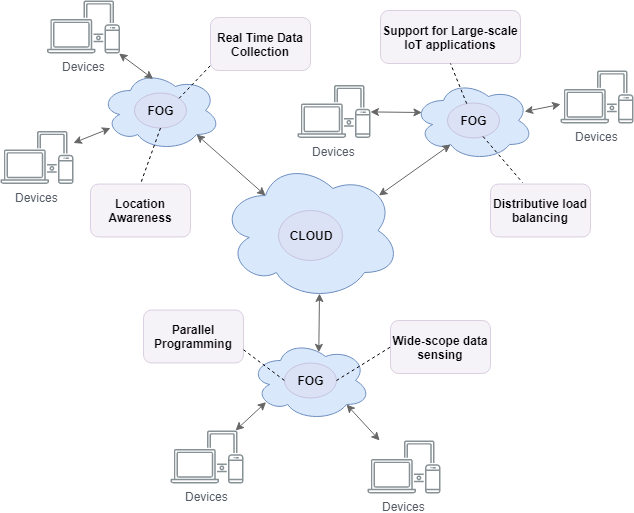}
    \caption{Fog-based IoT network.}
    \label{edge-fig}
\end{figure}

\subsection{Edge/Fog computing}
Edge or Fog computing is an alternative to cloud computing that can be used to offload the storage and computations from the IoT devices \cite{edge-root}. The difference is that cloud computing uses a server while fog computing uses a network edge or an edge device. It is an end-user device, located close to the IoT network~\cite{edge-promise}. It not only provides data but also processes data.  In fog computing, the end-user device requests the services and also handles the computing task. Cloud computing offloads the data management to the server while edge computing distributes the management load towards its edges. 

Offloading the computations from the IoT network is made possible with the use of cloudlets. A cloudlet is an edge device located between the server and the data source \cite{edge-promise}. It effectively handles computing tasks that include the processing, storage, caching, and load balancing of the data transported to and from the cloud. With the existence of this intermediary data processing area, the computation is offloaded and distributed to its edges. Figure \ref{edge-fig} shows a diagram that visualizes a fog-based IoT network and some of its notable features \cite{fog-app}. This computing design points out benefits and improvements to the standard cloud computing infrastructure\cite{edge-benefits, edge-arch}:

\begin{enumerate}[i.]
    \item \textit{Better offloading and reduced server strain} - Fog computing allows the network to offload the data processing to its cloudlets. Offloading reduces any resulting data traffic and server strain. Therefore, more users can now be managed more efficiently with edge computing. 
    
    \item \textit{Scalability through parallelism} - Fog computing allows scalability through parallelism and decentralization. It focuses on adding edge devices to the network. On the other hand, cloud computing scales a cloud server by increasing its size in terms of data capacity. 

\end{enumerate}

\subsection{Related Literature on Cloud and Fog Computing}
In \cite{multitier-fog}, the authors created a general scope of the potential of fog computing in terms of large scale data analysis. They proposed a multitier architecture for smart cities with the use of fog computing. Their paper opens a possibility for fog computing to drive the next wave of smart city applications. Urban sound classification is one of the intended sensing platforms for smart cities. By being able to create a sensory architecture to gather data around a large area, smart cities are made to be a more plausible implementation. However, their design provides an important detail that our paper aims to reinforce. Cloud computing still carries a critical role in enabling the fog. 

A similar design was discussed in \cite{big-data-smart}. They proposed a hierarchy of layers that combined fog and cloud technology to handle data from a larger scale. The network collects the data through the edge devices. Each device is grouped locally using cloudlets. The data then gets pre-processed from the multiple fog layers until it can be analyzed and stored globally within the cloud. Our paper takes the same stance as we aim to utilize the cloud as more than just a storage device but also a medium for more general computing. 

Both papers focused on a feature to benchmark the performance of their designs. \cite{multitier-fog} decided to use the number of jobs completed by each fog layer at given time intervals. On the other hand, \cite{big-data-smart} chose to test their design according to the accuracy of their classifier. Our papers differ as we decided on using power consumption, runtime, and latency. We selected these metrics to achieve a more in-depth investigation on the behaviour of the server and the end devices at different processing load configurations. Although the previously mentioned works were able to create a framework for a design that incorporates both cloud and fog computing, there are drawbacks in offloading processes within the network hierarchy. 

Moving computing and processing loads from the server to the edges require devices to have more processing power. As a result, the power consumption and application runtime among edges increases. Too much of this offloading might render a system undeployable. However, focusing on minimizing the power and runtime requirements from the devices will lead to server strains. Therefore, a proper load balance is needed within the network to make our system scalable. By focusing on these factors, we create a metric to base the feasibility of our proposed sensing platform. To test our framework, we needed an application that demanded a large amount of attention to power consumption, runtime, and server strain. Our investigation pointed us towards urban sound classification.

\subsection{Urban Sound Classification}
Urban sound classification is a sound sensing platform that distinguishes the types of sound that are present in a city-related area \cite{Urbansound}. Ideally, the sound data should be obtained by the devices in real-time. This constraint means that devices are recording sound almost all the time. As a result, an efficient configuration that minimizes power consumption and runtime must be in place to maximize the lifespan of the system. Also, the intention of this system is to be deployed and left to record sound from its assigned environment for long periods. Most sensing nodes are powered using batteries. This feature makes power efficiency crucial in maintaining the system. Also, the end devices are continuously transmitting data to the server. Without a proper balancing of computational loads, any server can easily be overloaded and rendered unable to function. Sound classification, in general, is composed of two stages; first, data collection and then, classification. In our implementation of the system, incoming data is first processed by the devices into a form that can be used by the neural network for classification. Then, the transformed data is fed by either the device or the server into the classifier for results. Within our implemented system, we trained the models to distinguish a selected pool of urban sounds.

\subsubsection{Feature Extraction}
Feature extraction is the process of obtaining related information from data \cite{mfcc}. However, data has become too large for conventional processing. Fortunately, even though data gets too big, it tends to be redundant. In that case, we can transform this type of data into a representative set of feature. Converting this data from its original form into a smaller set is called feature extraction. The following is a discussion of the sources of features that we used in our system.
 
\begin{enumerate}
\item \textit{MFCC and Mel-scale Spectrogram.}
Mel-frequency Cepstral Coefficient is a technique that extracts features in the cepstral domain \cite{mfcc}. The cepstral domain is a mathematical algorithm that obtains the envelope of the spectrum in the logarithm domain. First, the acoustic signal is converted by the function into a digital signal. This signal represents each level of the signal at every discrete time step. Next, the data is filtered, which maximizes the magnitude of the highest frequencies in comparison to its lower counterparts. This procedure is called pre-emphasis, which is performed to improve the overall signal-noise ratio of the signal.

 After pre-emphasis, the sound samples undergo a short-time Fourier Transform (STFT). In this procedure, the signal is framed and windowed by the STFT function. After removing any discontinuities from the resulting output of STFT, it undergoes Fast Fourier Transform (FFT). The result is a spectrogram which is a visualization of the signal in the frequency domain. Then, we apply a Mel-Scale filter to the signal, which yields a Mel-scale spectrogram. The features from this signal were obtained by relating the perceived frequency to actual frequency \cite{mel-scale}. 
 
 The ability to distinguish pitches and relate them to real-world sounds can help the neural network to even classify digital sounds from recordings or radio. The MFCC variation of this signal was still obtained to widen the feature scope of the classifier. Before obtaining the cepstral coefficients, the spectrum undergoes a logarithmic process. It is a process where we take the logarithm from the powers of each Mel frequency. Then, the resulting log powers undergo discrete cosine transform (DCT). DCT is a type of fast algorithm that transforms the domain of a set of signals from times to frequency \cite{dct}. The resulting frequency spectrum is the cepstral coefficients.

\item \textit{Chromagram.}
Chromagram is a technique that visualizes pitches as energy distributions. Also known as Harmonic Pitch Class Profile, chromagrams are spectrums that distinguish different pitch classes. There are multiple ways of obtaining the chroma spectrum. In this paper, Librosa uses the STFT approach. This approach starts with the sound input visualized in the frequency domain using STFT. Then the signal power is calculated using a logarithmic scale which results in frequency bins. These bins are eventually added together mathematically. Now expressed as energies, the values are warped onto one octave to fit the salience of 12 pitch classes \cite{chromagram}.
 
\item \textit{Spectral Contrast.}
Octave-Based Spectral Contrast is used to extract musical features by considering the difference between the spectral peaks and valleys in a spectrum. These peaks correspond to harmonic components, while the valleys are non-harmonic components such as noises. The computation process begins by expressing the sound input into the frequency domain via FFT. Then, we apply the Octave Scale Filters to the output, which will divide it into sub-bands. We then estimate the strength of spectral peaks, valleys and their differences in each sub-band. Next, we translate the resulting spectrum in Log domain. Finally, with the use of the Karhunen-Loeve transform, the features are mapped in orthogonal space \cite{contrast}.

\item \textit{Tonnetz.}
Tonnetz is a harmonic network representation of pitch intervals \cite{tonnetz}. The first stage is the Constant-Q spectral analysis which is a derived logarithmic frequency. These spectrum vectors are then framed and windowed into 12-bits. The resulting stream of bits is then transformed into Tonal Centroids in the 6-D space based on harmonic associations between pitches. The sequence of tonal centroid vectors is modulated using the overall rate of change of the smoothed tone and the distance between the vectors. Finally, we obtain the features from the resulting peaks in the signal. 
\end{enumerate}

\subsubsection{Neural Networks}
An artificial neural network is a computing system that is widely used by programs for classification \cite{neural-net}. It is a biological neural behaviour that has been observed and applied through computational algorithms. Neural networks come in different shapes and forms based on their intended learning flow. Each network models a general system that decides on an output based on its given input. A neural network can consist of three types of layers: input layers, output layers, and hidden layers. First, the input layer contains all the possible features or parameter values. Neural networks use these values for decision-making. Next, the output layer contains all the possible outcomes of the model given the possible inputs. Lastly, the hidden layer consists of a variable number of layers. These layers decide the different ways an input can take to reach an output. The network is trained programmatically using a given dataset.

The model is built by having the data simulated over a defined set of parameters and outcomes. By having the model observe the behaviour of the data during different instances, it can deduce patterns and relationships between inputs that eventually create the neural network \cite{neural-learning}. Neural networks are not guaranteed to be accurate. It can only be as precise as their training data. The training data can dictate all the possible relationships and behaviours that the neural network can observe. Therefore, the quality of the data(i.e. no noise or outlying information) determines the effectiveness of the network's training. Using a more complex classifier could limit us from moving the neural network to the device due to processing constraints. Therefore, we elected to settle with a more basic neural network to allow a more flexible system. 

\section{Methodology} \label{method}

\subsection{Overview}
The proposed system is composed of a cloud server connected to multiple end devices. Each device is programmed to communicate with the server wirelessly. The design starts by modelling the two standard data processing practices for IoT based classifiers; cloud and edge computing. Depending on the configuration, the sound data is either classified within the device or the server. Our focus on testing is on power and latency. Power consumption is one of the mentioned bottlenecks in terms of scaling a system. The sound classification was the selected application due to its high processing requirements that result in high demand in power from its sensing nodes or end devices. Using this type of application allowed us to point out any power usage differences between the two practices. Latency is another limiter for servers that attempt to scale up by adding more connected users. Servers tend to run into issues in managing incoming data once it reaches a certain threshold of connected devices or data traffic. To achieve a scalable design, we aim to investigate the feasibility of each data processing practice by simulating them under the proposed sound classification framework.  

\begin{figure}
    \centering
    \includegraphics[width=1\columnwidth]{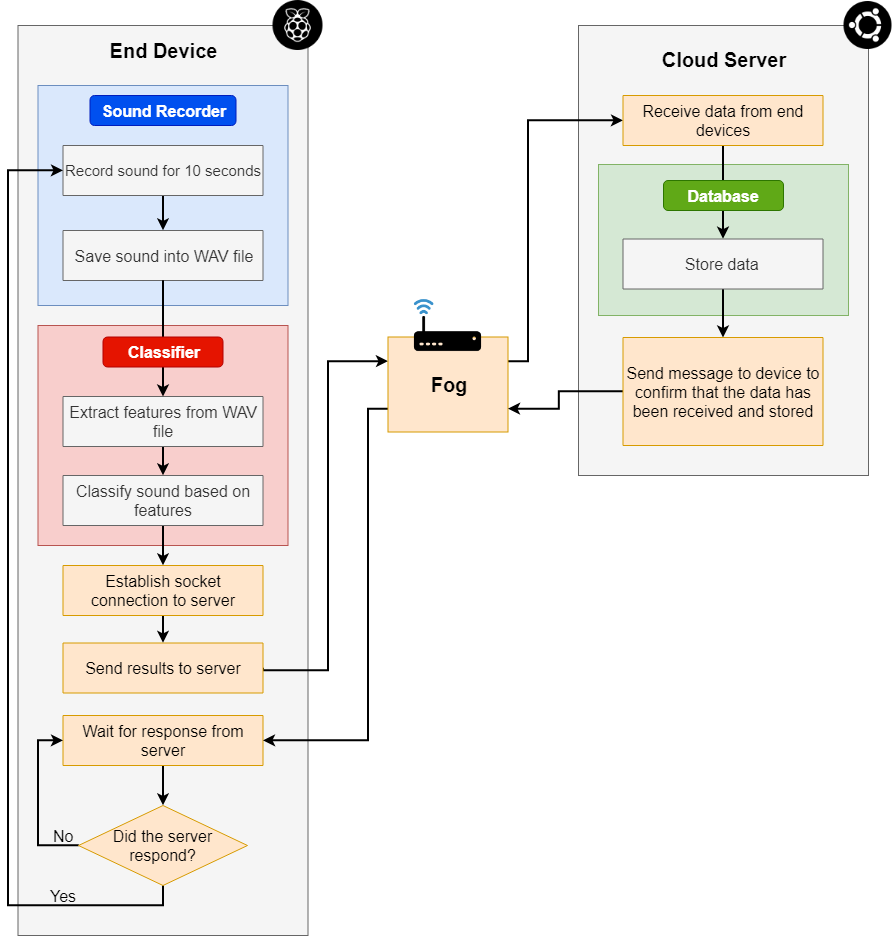}
    \caption{Edge/Fog-based setup logic flow.}
    \label{edge-flow}
\end{figure}

We based the first setup on edge computing. In this setup, the device does both the recording process as well as the classification. We labelled it as configuration A. The general flow starts similarly with configuration A where the end devices collect the sound data from the environment. However, instead of sending the sound data to the server for classification, the data is classified within the device. The results are then wirelessly sent to the server storage. Then, the server notifies the device via a message once the results have been received and stored. Figure \ref{edge-flow} shows the logical flow of the edge/fog setup or configuration A as implemented in our design. 

\begin{figure}
    \centering
    \includegraphics[width=1\columnwidth]{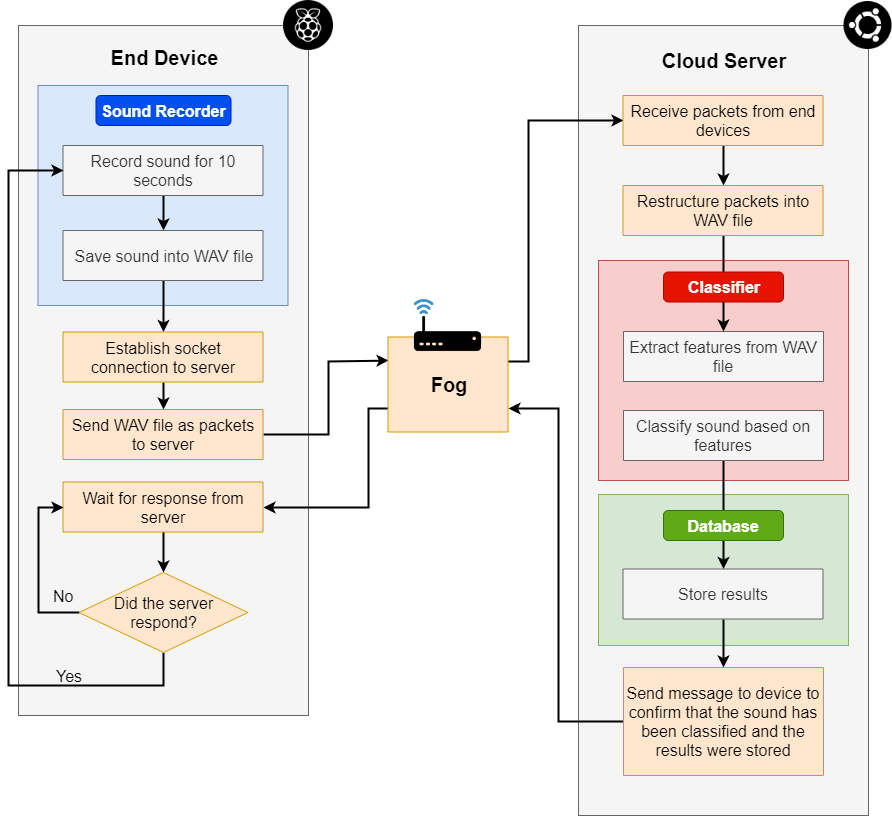}
    \caption{Cloud-based setup logic flow.}
    \label{cloud-flow}
\end{figure}

On the other hand, we based the second setup on cloud computing. We labelled it as configuration B for the conducted experiments. The general flow of this setup focuses all of the processing to the cloud. It starts with the edge devices, which is in charge of collecting the sound from the environment. Ideally, each device will be placed strategically around an area to get as much sound data as possible. Then, the end devices will send the sound data to the cloud server wirelessly. After receiving the data, the server converts these sound files into feature maps. Each feature map is used to classify the incoming sound. Then, the source device is notified by the server that it has successfully stored the classification results. Figure \ref{cloud-flow} shows the logical flow of the cloud-based setup or configuration B as implemented in our design. 

The purpose of starting with these two setups was to distinguish the differences in processing time and power consumption of the two represented computing practices in a controlled environment.   

\subsection{Experimental setup}
The experimental setup is composed of 4 major sections. Each aspect contributes to the previously discussed data processing practices and the implementation as a sound classifier. These sections are as follows; Design Specifications, Sound Recorder, Sound Classifier, and Testbed Setup.

\subsubsection{Design Specifications}
In terms of hardware used, both practices are the same. This decision was made to eliminate any extraneous variables on the experimental setup.  To represent the end devices of the network, using a Raspberry Pi 3 Model B. We used Pi's because they were easier to program and were capable of rapid prototyping. Each Pi 3 model B uses a Quad-core ARM Cortex A53, with a processing benchmark of 1.2 GHz. During experiments, multiple Pis are connected to the server using a star topology arrangement. Each Pi was pre-loaded with the default Raspbian-Jesse operating system for the end device implementations. As for the cloud server, the design makes use of a remote server that runs on an Ubuntu 16.04 operating system. For hardware specifications, the server used an Intel Xeon Processor E5-2640 that had 6 cores running at 2.50 GHz each. The Pis and the server contains Python with version 3.6 to cater to the software needed by the libraries in the design.  An STM32 NUCLEO-64 board with an attached X-NUCLEO-CCA02M1 expansion board to serve as a digital MEMS (Micro-Electrical-Mechanical System) microphone to record the sound. Using a digital microphone made it easier for the data that was read in by the Pi to be processed.    

\subsubsection{Sound Recorder}
The sound recording part of the implementation uses a digital microphone and a Python script programmed within the Pi. This script uses the microphone wired through a USB (Universal Serial Bus) on the Pi and a Python library called PySound. There was no need for any initial data processing because the microphone was digital. The microphone setup used for the STM32 NUCLEO-64 board and the X-NUCLEO-CCA02M1 were their default configurations (i.e.16 kHz sampling rate,  single-channel, and 16-bit resolution). The script specifies a variable recording length to help in the classifier later on.  It calls the PyAudio library to initialize the recording by creating an audio stream that uses the specifications of the digital microphone.

During recording, the Pi reads in frames of data from the microphone. The selected frame size was 4096 bytes per frame. Each frame is appended to a list by the script. Upon reaching the recording length, the library closes the stream and a file handler stores the data into a WAV file. Afterwards, depending on the data processing practice, the sound file is either sent to the server via Socket Transmission Control Protocol (TCP) or kept within the Pi for classifying.

\subsubsection{Sound Classifier}
The sound classifier implementation consists of 3 sections: the dataset, the feature extraction process, and the learning architecture. 

\begin{enumerate}
\item \textit{Dataset.}
The classifier used in the experiment uses an UrbanSound8K dataset \cite{Urbansound}. This dataset is a group of sound files compiled from different sound types in an urban setting. The original UrbanSound dataset is a collection of 1302 full-length recordings. Each sound,  labelled according to the sound occurrence and salience annotations,  varied from a duration of 1-2 s to over 30 s. The 8K version of the dataset was a modified version of the original. This version splits the 1302 recordings into 4-second clips resulting in 8732 sound excerps. Listening tests were conducted by Salamon \cite{Urbansound} to find out the most optimal sound format to achieve the best identification accuracy. The results indicated that 4 seconds was the best clip duration which yielded an accuracy of 82\%. The UrbanSound8K dataset consists of 10 low-level classes: air conditioner, car horn, children playing, dog barking, drilling, engine idling, gunshot, jackhammer, siren and street music. We based our classifier on the design advised by Salamon \cite{Urbansound}. Also, in terms of the dataset, he arranged the sound clips in 10 folders. Each folder contained around 850 clips upon average. We will be using these folders to train the model outlined in this paper.

\item \textit{Feature Extraction.}
To extract these features, the script imports a sound processing library called Librosa. Librosa loads in the dataset and converts them into feature maps. The conversion is carried out using feature extraction methods provided by Librosa. Also, this paper consists of 5 feature extraction sources: MFCC, Chromagram, Mel-scaled Spectrogram, Spectral Contrast, and Tonnetz. With the use of these sources, we created a program that transformed all the sounds related to the classifier into sound maps. 

\item \textit{Neural Network.}
After extracting the features, the scripts compile them into a sound map and then fed into the neural network. The neural network uses Tensorflow to construct its framework. Our classifier uses a basic neural network framework that has 2 hidden layers. Also, each one contains 280 and 300 nodes, respectively. We used Tensorflow and three folders from the Urbansound8K dataset to train the neural network. Each folder consists of 870 clips that were 4 seconds long as described by Salamon \cite{Urbansound}. Also, we resampled all the clips as 22050 Hz sound files for feature extraction. We then configured the training scripts to have variable training epochs and learning rate to obtain the highest accuracy. We split the dataset 70-30, where 70\% of the sound files were used to train the classifier. The remaining 30\% was used to test and verify the classifier afterwards. After training and testing the classifier under multiple iterations, the best accuracy was 85\%. This result came from a training epoch of 5000 and a learning rate of 0.1.  
\end{enumerate}

\subsubsection{Testbed Setup}
Three tests were conducted to obtain the three metrics that are the focus of this paper. The tests were to measure the end device power consumption, application runtime and the server latency. The testbed takes these three tests and breaks them into two sections: the power consumption and the runtime test, and the test for latency. For the first test, we executed a script that samples a 10 second sound clip through the microphone. After finishing the recording, the program saves the sound clip as a WAV file. Next is another script that reads in the WAV file and extracts its features. An urban sound classifier model defined in the neural network section then uses these features. Once the model returns the results, the program then exits.

Firstly, the testbed measures the power consumption of the end devices by using the Monsoon power monitor. This test allows the observation of any significant differences between the two practices in terms of their effective execution. This power monitor will be used to directly power the Pis to measure any changes in power consumption while running the program. By taking note of when these changes occur and aligning them with the recorded timestamps, the testbed can be used to map out the overall power usage of the device during testing. Next, the testbed measures the runtime by adding timestamps during every major procedure within the program. In measuring the total runtime of the application, each end device will wait for a confirmation from the server that the results have been successfully obtained and stored by the classifier. 

The next test takes each implementation and measures its latency by measuring the times it takes for the Pi to transmit all of the data to the server. Each Pi will be sending packets to the server at the same time. A round-robin type of scheduling is used to manage each Pi to obtain each of the data they are transmitting sequentially. Each transmission is timestamped to check when a Pi starts a sending and when it ends. This design allows us to measure the latency of each Pi and pool them together to obtain an average value representative of each configuration.

\section{Testing and Evaluation} \label{eval}

The purpose of the tests was to determine the most optimal configuration of our design. By considering the different types of computing architectures discussed in this paper, we attempted to create the best iterations of the proposed system. 

Initially, our test takes two configurations. The first configuration has the classification done on the edge while the other has the classifier inside the server. These configurations were labelled configurations A and B, respectively. Configuration A was set up by placing the classifier in the Raspberry pi to be executed in the end device. First, the operation flow starts with the microphone recording the sound for 10 seconds. Next, the program will save the sound data into a file that is read by the classifier script. The classifier code will then extract a feature map from it. Then, our model will classify the results of the extraction. Lastly, the results of the classifier are sent by the devices to be stored by the server. 

Configuration B was set up by placing the classifier in the server to be executed in the cloud. The recording is the same with configuration A. However, instead of classifying the sound file in the Pi, it will be sent wirelessly via packets to the server. Using wireless socket communication, the Raspberry Pi will establish a connection to the server. Once a connection was established, the Pi will send the sound file to the server as packets. The server will then reconstruct this information into a workable file. Finally, the program will classify the sound and store the results into the database.

\begin{figure}
    \centering
    \includegraphics[width=1\columnwidth]{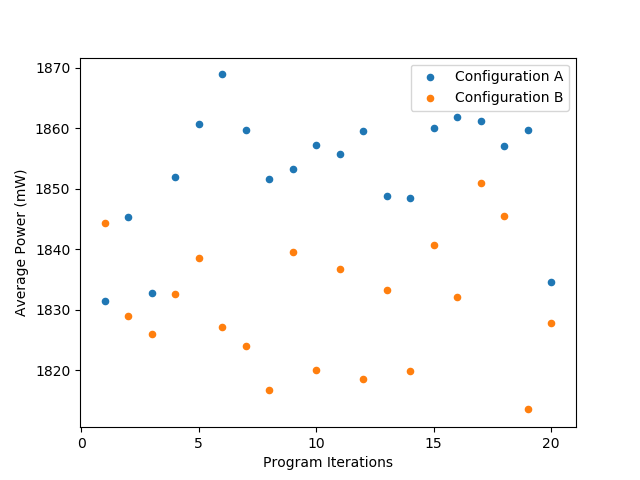}
    \caption{Power consumption comparison between configuration A and B over 20 iterations.}
    \label{pow-locserv}
\end{figure}

Both configurations were executed for 20 iterations. The first metric measured was the power consumption of each iteration. Figure \ref{pow-locserv} shows a comparison between configurations A and B in terms of their respective iterations. It can be observed through the graph that most of the points from configuration A is higher than that of configuration B. The results yielded an average power consumption of 1852.00 mW for configuration A and 1830.54 mW for B. Configuration B has a lower power consumption than A by a value of 21.46 mW. However, this value is not large enough to be considered a significant difference. Therefore in terms of power consumption, configuration A and configuration B are relatively the same.

We also measured the runtime of the program for 20 iterations. To make sure that the observations between the runtime of the two configurations are more distinct, we broke them down into two runtime measurements: the recording phase and the classification phase. As mentioned in the setup, the end device will only record the runtime once the server confirms that all the server-side processes (i.e. Classification for configuration B or storage for configuration A) are complete. 

Over the same 20 iterations, we also measured the runtime of the program. To make sure that the observations between the runtime of the two configurations are more distinct, we broke them down into two runtime measurements: the recording phase and the classification phase. As mentioned in the setup, the end device will only record the runtime once the server confirms that all the server-side processes (i.e. Classification for configuration B or storage for configuration A) are complete. 

\begin{figure}[t!] 
    \centering
    \includegraphics[width=1\columnwidth]{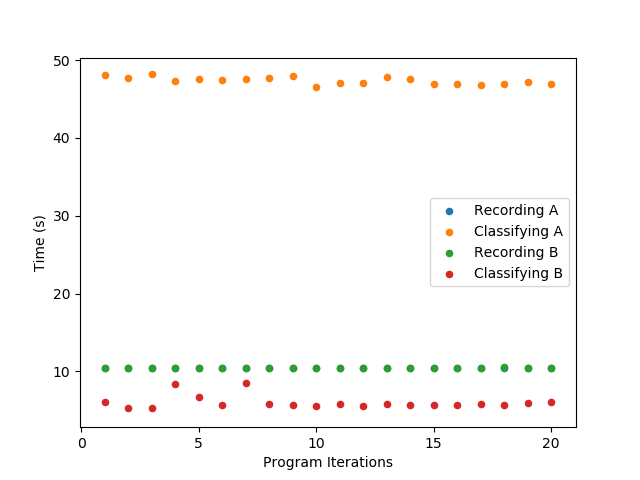}
    \caption{Runtime comparison between configuration A and B over 20 iterations.}
    \label{run-locserv}
\end{figure}

Figure \ref{run-locserv} shows a  graphical representation of our runtime measurements for each phase. We observed that the recording phase for both configurations is the same. However, the time it takes to complete the classification phase has configuration B taking 6.02 seconds while configuration A takes 47.36 seconds. This significant difference shows that configuration B is better than configuration A in terms of runtime by being 41.31 seconds faster overall. Overall, we calculated an average runtime of 57.77 and 6.42 seconds for configurations A and B, respectively.  

\begin{figure}
    \centering
    \includegraphics[width=1\columnwidth]{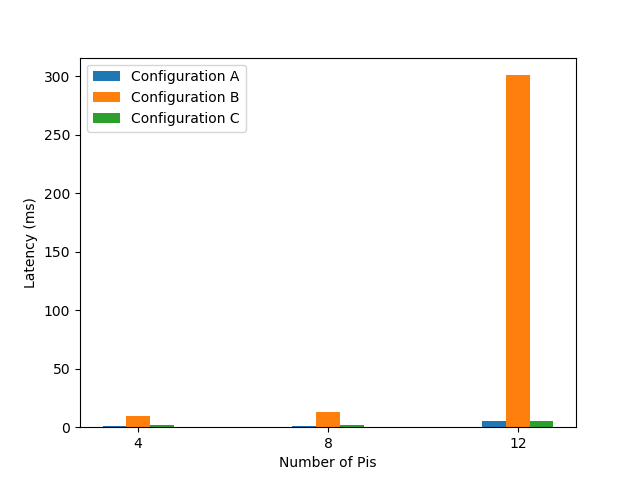}
    \caption{Latency comparison between configurations A, B, and C.}
    \label{lat-locserv}
\end{figure}

The next test was to investigate the latency of each configuration. Each configuration was tested by having multiple Pis to transmit their data to the server. The server receives the data, one Pi at a time through round-robin scheduling. We intended to deploy the design to support growing networks. Therefore, the test was conducted with 4, 8, and then 12 Pis to simulate the effects of a growing network to the latency of each configuration. Figure \ref{lat-locserv} shows that results of the latency test. We observed that configuration B significantly slows down as the number of Pis increases. This observation exposes the inability of the configuration to manage growing networks. On the other hand, configuration A shows that it can manage larger systems through its steady graph. However, as we recall the previous test, configuration A is significantly slower in terms of runtime than B.

\begin{table}[t!]
\centering
\setlength\tabcolsep{0.02cm}
\begin{tabular}{|c|c|c|}
\hline
 & \multicolumn{2}{c|}{\textbf{Metrics}} \\ \hline
\textbf{Configuration} &  \multicolumn{1}{c|}{\begin{tabular}[x]{@{}c@{}}Average Power \\ Consumption (mW) \end{tabular}} & \multicolumn{1}{c|}{Average Runtime (s)} \\ \hline
A &  1852.00 & 57.77  \\ \hline
B & 1830.54  & 16.42  \\ \hline
C &  1786.86 & 53.02 \\ \hline
\end{tabular}
\caption{Testing results for power consumption and runtime.}
\label{table-pow}
\end{table}

\begin{table}[t!]
\centering
\begin{tabular}{|c|c|c|c|}
\hline
 & \multicolumn{3}{c|}{\textbf{Average Latency (ms)}} \\ \hline
\textbf{Configuration} & \multicolumn{1}{c|}{4} & \multicolumn{1}{c|}{8} & \multicolumn{1}{c|}{12} \\ \hline
A &  0.6 & 1.0 & 5.4 \\ \hline
B & 9.7 & 13.0 & 300.7 \\ \hline
C & 1.7 & 2.2 & 5.5\\ \hline
\end{tabular}
\caption{Testing results for latency.}
\label{table-lat}
\end{table}

To investigate further, we tried to take the slowest section of configuration A and offload some of its parts to the server. As a result, we created a new configuration that had the feature extraction conducted in the Pi and the actual classification in the server. So instead of sending a whole sound file to the server, the Pi sends a feature map instead. In theory, this could improve some aspects of the design by requiring less power and reducing latency by transmitting fewer data. 

The new configuration, labelled as C, was tested similar to A and B.  Table \ref{table-pow} shows a summary of the power and runtime test comparing the results. Configuration C yielded an average power consumption of 1786.86 mW being lower than both A and B. Also, the average runtime that we measured was 53.02 seconds.  This measurement shows a potential drawback to this configuration by being only slightly faster than A, but still significantly slower than B. Lastly, we measured its latency. Table \ref{table-lat} shows a summary of the latency test comparing the results. We observed that configuration C is steadier than B in terms of latency increase over network growth. Also, its values and trend appear to be the same as A.  

As observed from the yielded values, C is the best in terms of power consumption. In terms of runtime, B is the best leaving the other two significantly slower. However, in the last test, C and A show their ability to handle growing networks by maintaining relatively low latencies at higher network traffic while B does not.  Though B can be faster, it is not scalable since it cannot handle an increase in sensing nodes. Therefore, run time does not indicate any advantages of B over the other configuration. This puts more weight towards power consumption and latency than runtime. Each configuration is ranked. To have a more systematic way of determining the best among the 3 configurations, each rank is assigned a specific number of points, 1 for third, 2 for second and 3 for first. The configuration that has the highest number of points is arguably the best. For the runtime and power consumption, the scoring is based on the obtained values. While for the latency, the scoring was based on the effectiveness of the configuration as the number of Pis increased. Table~\ref{table-rank} shows the scoring of each configuration with respect to the data collected. By consulting the data and the results, the best configuration is C which is a balanced combination of both edge and cloud computing architectures.




\begin{table}[t!]
\setlength\tabcolsep{0.08cm}
\centering
\begin{tabular}{|c|c|c|c|c|}
\hline
 & \multicolumn{3}{c|}{\textbf{Scoring}} & \\ \hline
\textbf{Configuration} & \begin{tabular}[x]{@{}c@{}}Power \\ consumption \end{tabular} & \begin{tabular}[x]{@{}c@{}}Runtime \end{tabular} & \begin{tabular}[x]{@{}c@{}}Latency \end{tabular} & \multicolumn{1}{c|}{Tally} \\ \hline
A & 1 & 1 & 3 & \textbf{5}\\ \hline
B & 2 & 3 & 1 & \textbf{6}\\ \hline
C & 3 & 2 & 2 & \textbf{7}\\ \hline
\end{tabular}
\caption{Configuration scoring.}
\label{table-rank}
\end{table}

\section{Conclusion} \label{conc}
This paper examines  two types of computing; Edge/Fog computing and Cloud computing. Each type focuses on a specific aspect of the IoT network. Tests were conducted to investigate on the optimization techniques to create an efficient system in terms of minimizing both end device power consumption and server-side strains for sound classification for citywide deployment. 

There were 3 configurations used in the testing. The first configuration took the classifier and executed it within the end device. The second configuration moved the classifier to the server. The third configuration took apart the classifier into two sections, placing the feature extraction in the end device and the classification in the server. First, the performance of the classifier was tested. The power consumption and runtime of the system under each configuration was observed. It was determined that in terms of power consumption, the third classification was the best by requiring less power to transmit the data that was significantly reduced in size. 

For run time, the second configuration was the best significantly due to the processing power of the server in comparison to the end device. Next, the scalability of the system was measured by testing its latency as the number of end devices were increased. The results showed that the first and third configurations were capable of scaling. On the other hand, the latency of the second configuration significantly increased as the number of end devices increased. Lastly, each configuration was ranked based on the results of the test. The scoring showed that the third configuration was the best overall. Therefore, it shows how focusing on a single type of computing could result in a limited design. 

The future of IoT-applications in this growing industry is by finding optimal configurations and distributions of work within a system. Not focusing on a single archetype can prove to be beneficial in creating a more power efficient and load balanced IoT-based system.  

\section*{References}
\bibliography{edgebib}

\begin{thebibliography}{10}
\expandafter\ifx\csname url\endcsname\relax
  \def\url#1{\texttt{#1}}\fi
\expandafter\ifx\csname urlprefix\endcsname\relax\def\urlprefix{URL }\fi
\expandafter\ifx\csname href\endcsname\relax
  \def\href#1#2{#2} \def\path#1{#1}\fi

\bibitem{iot-def}
R.~{Muñoz}, R.~{Vilalta}, N.~{Yoshikane}, R.~{Casellas}, R.~{Martínez},
  T.~{Tsuritani}, I.~{Morita}, Integration of iot, transport sdn, and
  edge/cloud computing for dynamic distribution of iot analytics and efficient
  use of network resources, Journal of Lightwave Technology 36~(7) (2018)
  1420--1428.
\newblock \href {http://dx.doi.org/10.1109/JLT.2018.2800660}
  {\path{doi:10.1109/JLT.2018.2800660}}.

\bibitem{Al-Fuqaha}
A.~{Al-Fuqaha}, M.~{Guizani}, M.~{Mohammadi}, M.~{Aledhari}, M.~{Ayyash},
  Internet of things: A survey on enabling technologies, protocols, and
  applications, IEEE Communications Surveys Tutorials 17~(4) (2015) 2347--2376.
\newblock \href {http://dx.doi.org/10.1109/COMST.2015.2444095}
  {\path{doi:10.1109/COMST.2015.2444095}}.

\bibitem{iot-cloud}
M.~{Shiraz}, A.~{Gani}, R.~H. {Khokhar}, R.~{Buyya}, A review on distributed
  application processing frameworks in smart mobile devices for mobile cloud
  computing, IEEE Communications Surveys Tutorials 15~(3) (2013) 1294--1313.
\newblock \href {http://dx.doi.org/10.1109/SURV.2012.111412.00045}
  {\path{doi:10.1109/SURV.2012.111412.00045}}.

\bibitem{Takabi}
H.~{Takabi}, J.~B.~D. {Joshi}, G.~{Ahn}, Security and privacy challenges in
  cloud computing environments, IEEE Security Privacy 8~(6) (2010) 24--31.
\newblock \href {http://dx.doi.org/10.1109/MSP.2010.186}
  {\path{doi:10.1109/MSP.2010.186}}.

\bibitem{fog-app}
J.~{Ni}, K.~{Zhang}, X.~{Lin}, X.~S. {Shen}, Securing fog computing for
  internet of things applications: Challenges and solutions, IEEE
  Communications Surveys Tutorials 20~(1) (2018) 601--628.
\newblock \href {http://dx.doi.org/10.1109/COMST.2017.2762345}
  {\path{doi:10.1109/COMST.2017.2762345}}.

\bibitem{road-smart}
J.~{Rivas}, R.~{Wunderlich}, S.~J. {Heinen}, Road vibrations as a source to
  detect the presence and speed of vehicles, IEEE Sensors Journal 17~(2) (2017)
  377--385.
\newblock \href {http://dx.doi.org/10.1109/JSEN.2016.2628858}
  {\path{doi:10.1109/JSEN.2016.2628858}}.

\bibitem{cloud-sec}
P.~{Mell}, What's special about cloud security?, IT Professional 14~(4) (2012)
  6--8.
\newblock \href {http://dx.doi.org/10.1109/MITP.2012.84}
  {\path{doi:10.1109/MITP.2012.84}}.

\bibitem{cloud-ben}
D.~{Zhe}, W.~{Qinghong}, S.~{Naizheng}, Z.~{Yuhan}, Study on data security
  policy based on cloud storage, in: 2017 ieee 3rd international conference on
  big data security on cloud (bigdatasecurity), ieee international conference
  on high performance and smart computing (hpsc), and ieee international
  conference on intelligent data and security (ids), 2017, pp. 145--149.
\newblock \href {http://dx.doi.org/10.1109/BigDataSecurity.2017.12}
  {\path{doi:10.1109/BigDataSecurity.2017.12}}.

\bibitem{cloud-edge-comp}
Y.~{Zhou}, D.~{Zhang}, N.~{Xiong}, Post-cloud computing paradigms: a survey and
  comparison, Tsinghua Science and Technology 22~(6) (2017) 714--732.
\newblock \href {http://dx.doi.org/10.23919/TST.2017.8195353}
  {\path{doi:10.23919/TST.2017.8195353}}.

\bibitem{edge-root}
M.~{Gusev}, S.~{Dustdar}, Going back to the roots—the evolution of edge
  computing, an iot perspective, IEEE Internet Computing 22~(2) (2018) 5--15.
\newblock \href {http://dx.doi.org/10.1109/MIC.2018.022021657}
  {\path{doi:10.1109/MIC.2018.022021657}}.

\bibitem{edge-promise}
W.~{Shi}, S.~{Dustdar}, The promise of edge computing, Computer 49~(5) (2016)
  78--81.
\newblock \href {http://dx.doi.org/10.1109/MC.2016.145}
  {\path{doi:10.1109/MC.2016.145}}.

\bibitem{edge-benefits}
A.~C. {Baktir}, A.~{Ozgovde}, C.~{Ersoy}, How can edge computing benefit from
  software-defined networking: A survey, use cases, and future directions, IEEE
  Communications Surveys Tutorials 19~(4) (2017) 2359--2391.
\newblock \href {http://dx.doi.org/10.1109/COMST.2017.2717482}
  {\path{doi:10.1109/COMST.2017.2717482}}.

\bibitem{edge-arch}
M.~{Marjanovic}, A.~{Antonic}, I.~P. {Žarko}, Edge computing architecture for
  mobile crowdsensing, IEEE Access 6 (2018) 10662--10674.
\newblock \href {http://dx.doi.org/10.1109/ACCESS.2018.2799707}
  {\path{doi:10.1109/ACCESS.2018.2799707}}.

\bibitem{multitier-fog}
J.~{He}, J.~{Wei}, K.~{Chen}, Z.~{Tang}, Y.~{Zhou}, Y.~{Zhang}, Multitier fog
  computing with large-scale iot data analytics for smart cities, IEEE Internet
  of Things Journal 5~(2) (2018) 677--686.
\newblock \href {http://dx.doi.org/10.1109/JIOT.2017.2724845}
  {\path{doi:10.1109/JIOT.2017.2724845}}.

\bibitem{big-data-smart}
B.~{Tang}, Z.~{Chen}, G.~{Hefferman}, S.~{Pei}, T.~{Wei}, H.~{He}, Q.~{Yang},
  Incorporating intelligence in fog computing for big data analysis in smart
  cities, IEEE Transactions on Industrial Informatics 13~(5) (2017) 2140--2150.
\newblock \href {http://dx.doi.org/10.1109/TII.2017.2679740}
  {\path{doi:10.1109/TII.2017.2679740}}.

\bibitem{Urbansound}
J.~Salamon, C.~Jacoby, J.~P. Bello, A dataset and taxonomy for urban sound
  research, in: 22nd {ACM} International Conference on Multimedia (ACM-MM'14),
  Orlando, FL, USA, 2014, pp. 1041--1044.

\bibitem{mfcc}
L.~{Shi}, I.~{Ahmad}, Y.~{He}, K.~{Chang}, Hidden markov model based drone
  sound recognition using mfcc technique in practical noisy environments,
  Journal of Communications and Networks 20~(5) (2018) 509--518.
\newblock \href {http://dx.doi.org/10.1109/JCN.2018.000075}
  {\path{doi:10.1109/JCN.2018.000075}}.

\bibitem{mel-scale}
S.~{Umesh}, L.~{Cohen}, D.~{Nelson}, Frequency warping and the mel scale, IEEE
  Signal Processing Letters 9~(3) (2002) 104--107.
\newblock \href {http://dx.doi.org/10.1109/97.995829}
  {\path{doi:10.1109/97.995829}}.

\bibitem{dct}
E.~{Imam}, M.~E.~M. {Ahmed}, G.~{Abdalla}, Design and implementation of
  discrete cosine transform algorithm on fpga device, in: 2016 Conference of
  Basic Sciences and Engineering Studies (SGCAC), 2016, pp. 13--18.
\newblock \href {http://dx.doi.org/10.1109/SGCAC.2016.7457999}
  {\path{doi:10.1109/SGCAC.2016.7457999}}.

\bibitem{chromagram}
Y.~{Ni}, M.~{McVicar}, R.~{Santos-Rodriguez}, T.~{De Bie}, An end-to-end
  machine learning system for harmonic analysis of music, IEEE Transactions on
  Audio, Speech, and Language Processing 20~(6) (2012) 1771--1783.
\newblock \href {http://dx.doi.org/10.1109/TASL.2012.2188516}
  {\path{doi:10.1109/TASL.2012.2188516}}.

\bibitem{contrast}
D.~{Jiang}, H.~{Zhang}, J.~{Tao}, L.~{Lu}, L.~{Cai}, Music type classification
  by spectral contrast feature, in: Proceedings. IEEE International Conference
  on Multimedia and Expo, Vol.~1, 2002, pp. 113--116 vol.1.
\newblock \href {http://dx.doi.org/10.1109/ICME.2002.1035731}
  {\path{doi:10.1109/ICME.2002.1035731}}.

\bibitem{tonnetz}
E.~J. {Humphrey}, T.~{Cho}, J.~P. {Bello}, Learning a robust tonnetz-space
  transform for automatic chord recognition, in: 2012 IEEE International
  Conference on Acoustics, Speech and Signal Processing (ICASSP), 2012, pp.
  453--456.
\newblock \href {http://dx.doi.org/10.1109/ICASSP.2012.6287914}
  {\path{doi:10.1109/ICASSP.2012.6287914}}.

\bibitem{neural-net}
G.~P. {Zhang}, Neural networks for classification: a survey, IEEE Transactions
  on Systems, Man, and Cybernetics, Part C (Applications and Reviews) 30~(4)
  (2000) 451--462.
\newblock \href {http://dx.doi.org/10.1109/5326.897072}
  {\path{doi:10.1109/5326.897072}}.

\bibitem{neural-learning}
S.~{Marinai}, M.~{Gori}, G.~{Soda}, Artificial neural networks for document
  analysis and recognition, IEEE Transactions on Pattern Analysis and Machine
  Intelligence 27~(1) (2005) 23--35.
\newblock \href {http://dx.doi.org/10.1109/TPAMI.2005.4}
  {\path{doi:10.1109/TPAMI.2005.4}}.

\end{thebibliography}
\end{document}